\documentclass[
aps,amsmath,amssymb,reprint, double column,prl, showkeys,longbibliography]{revtex4-2}

\usepackage{graphicx,epsfig}
\usepackage{amssymb}
\usepackage{amsmath}
\usepackage{bm}
\usepackage{textcomp}
\usepackage{color}
\usepackage{float}
\usepackage{comment}
\usepackage{soul}
\usepackage{xcolor}

\begin{document}

\setcounter{page}{1}

\title[]{Engineering Quantum Criticality in the Integer Quantum Hall Regime through a Screening Layer}
\author{C.~T.~Tai}
\thanks{These authors contributed equally to this work.}
\author{P.~T.~Madathil}
\thanks{These authors contributed equally to this work.}
\author{A.~Gupta}
\author{L.~N.~Pfeiffer}
\author{K.~W.~Baldwin}
\author{M.~Shayegan}

\affiliation{Department of Electrical and Computer Engineering, Princeton University,
Princeton, New Jersey 08544, USA}

\date{\today}

\begin{abstract}
Disorder-induced localization of electrons and electron-electron interaction are among the most fundamental problems in condensed matter physics. In two-dimensional electron systems, extensive studies have led to the emergence of a scaling picture, characterized by a set of universal critical exponents that govern the transitions between the integer quantum Hall plateaus. From the temperature dependence of the plateau-to-plateau transitions, experiments primarily report $\kappa \simeq 0.42$, implying a dynamic exponent $z=1$, consistent with a theoretical picture where electrons have a long-range ($1/r$) interaction. Theory also predicts that $z$ = 2 for short-range electron interaction, but an experimental verification has remained elusive. Here, we directly probe the influence of Coulomb interaction on these transitions using a bilayer electron system confined to a GaAs double quantum well device. The two layers are in close proximity, with an interlayer distance approximately equal to the magnetic length at the relevant magnetic fields. By tuning the electron density in the top layer, we access both $insulating$ and $metallic$ phases of the electrons in this layer as a function of magnetic field, allowing \textit{in-situ} control of the $unscreened$ and $screened$ interaction strengths in the bottom layer as it goes through its plateau-to-plateau transitions. In the unscreened case, we measure $\kappa \simeq 0.42$ consistent with the widely reported value. More importantly, when screening is introduced, $\kappa$ is reduced to $\simeq 0.22$, implying $z$ = 2. Our results provide direct experimental evidence for the role of electron-electron interaction in determining critical behavior in the quantum Hall regime, and demonstrate screening as a powerful tuning parameter for engineering quantum criticality.
\end{abstract}

\maketitle 

\begin{figure}[t]
\centering
\includegraphics[width=1.0\columnwidth]{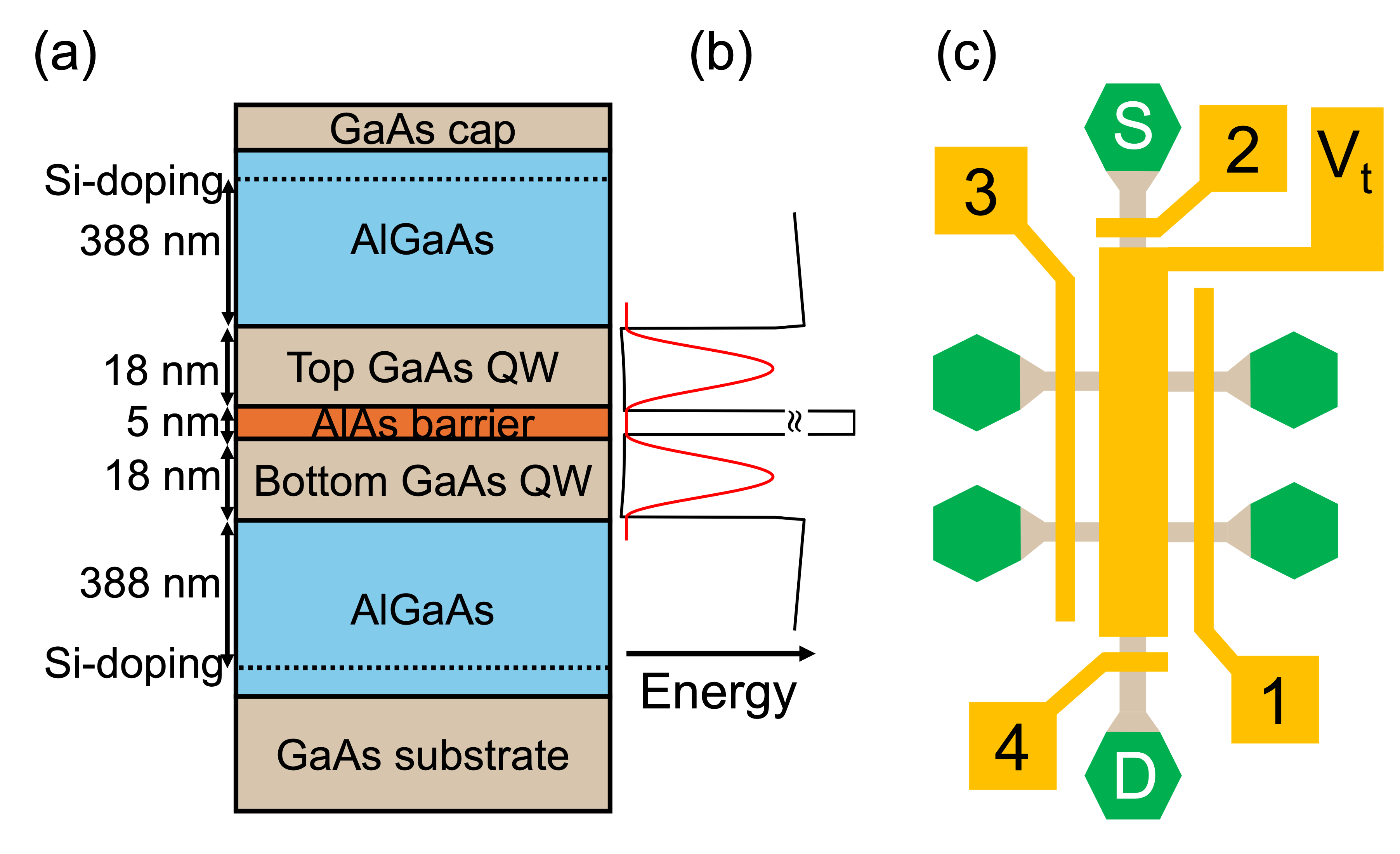}
    \caption{ (a) Schematic of the sample structure, consisting of two 18-nm-wide GaAs QWs separated by a 5-nm-wide AlAs barrier. (b) Charge distribution (red) and potential energy (black), calculated by solving the Schrödinger-Poisson equations self-consistently. (c) Schematic of the device structure. The gray area represents the Hall bar mesa, and the green hexagonal areas are the ohmic contacts which are used as voltage probes, or for injecting current through source (S) and drain (D). Finger gates 1-4 are used to deplete the top layer so that magnetotransport coefficients of the bottom layer can be measured. The global top gate ($V_{t}$) allows changing the top-layer electron density.}
\label{fig:1}
\end{figure}

The integer quantum Hall effect (IQHE)~\cite{klitzing1980prl} stands as a cornerstone of clean two-dimensional (2D) electron systems, underpinning fundamental advances in physics ranging from topology~\cite{thouless1982prl,halperin1982prb,avron1983prl}, precision metrology,~\cite{Poirier2009EuroPhysicalJournal,tzalenchuk2010natnano,Stock2019metrologia} to quantum phase transitions and criticality~\cite{pruisken1984nuclphysb, wei1986prb, Chalker1988jpc, Pruisken1988PRL, wei1988prl, Jain1990prl, Koch1991PRL, Koch1991PRB, kivelson1992prb, Wei1992PRB, Huo1993PRL, Engel1993prl, Aleiner1994prb, Wei1994PRB, Huckestein1995rmp, Lee1996PRL, Sondhi1997rmp, Gusev1998PRB, Huckestein1999prl, Hohls2002PRL, Li2005PRL, Li2009PRL, amoo2014JPCM, hui2019prb, arapov2019LTP, kumar2022prr}. In its simplest description, the IQHE manifests as a consequence of Landau quantization of the energy levels of non-interacting electrons in the presence of an external magnetic field. This leads to a characteristic fingerprint in electrical transport measurements, namely, the presence of quantized Hall resistance ($R_{xy}$) plateaus accompanied by vanishing longitudinal resistance ($R_{xx}$) at integer filling factors ($\nu$). Between adjacent quantized plateaus, the system undergoes a continuous quantum phase transition where extended states emerge that percolate throughout the sample, leading to a finite conductivity. These plateau-to-plateau transitions provide an ideal platform for studying quantum critical phenomena, particularly, the \textit{universality} of non-interacting and strongly-interacting systems, in the presence of disorder~\cite{pruisken1984nuclphysb, wei1986prb, Chalker1988jpc, Pruisken1988PRL, wei1988prl, Jain1990prl, Koch1991PRL, Koch1991PRB, kivelson1992prb, Wei1992PRB, Huo1993PRL, Engel1993prl, Aleiner1994prb, Wei1994PRB, Huckestein1995rmp, Lee1996PRL, Sondhi1997rmp, Gusev1998PRB, Huckestein1999prl, Hohls2002PRL, Li2005PRL, Li2009PRL, amoo2014JPCM, hui2019prb, arapov2019LTP, kumar2022prr}.\\
\indent Two critical exponents emerge at quantum Hall plateau transitions. The first is associated with the divergence of the localization length ($\xi$). At zero temperature, $\xi$ diverges at a single critical energy ($E_{c}$) according to the power law $ \xi \sim |E - E_{c}|^{-\gamma}$ with a universal critical exponent $\gamma \simeq 2.4$~\cite{Pruisken1988PRL,Huo1993PRL, Aleiner1994prb,Huckestein1995rmp}. The second is the dynamic exponent ($z$), which characterizes the divergence of the temporal correlation length ($\xi_{\tau}$), given by $\xi_{\tau}\sim \xi^z$~\cite{Sondhi1997rmp}. Experimentally, these exponents can be extracted from the temperature dependence of $R_{xy}$ and $R_{xx}$. Specifically, the derivative of $R_{xy}$ (with respect to the magnetic field $B$) at the critical magnetic field ($B_{c}$) and the inverse of the half-width of $R_{xx}$ between two successive quantum Hall states both diverge according to the power law $T^{-\kappa}$~\cite{Pruisken1988PRL}. $\kappa$ is linked to $\gamma$ and $z$ through the relation $\kappa = 1/\gamma z$~\cite{Huckestein1995rmp,Sondhi1997rmp,Koch1991PRL,Aleiner1994prb,Koch1991PRB,Engel1993prl,Wei1994PRB,Li2005PRL,Li2009PRL}. Experiments primarily report $\gamma \simeq 2.4$, and more importantly, $\kappa \simeq 0.42$, implying $z = 1$ for the IQH transitions~\cite{wei1988prl,Koch1991PRL,Koch1991PRB,Wei1992PRB,Engel1993prl,Aleiner1994prb,Wei1994PRB,Gusev1998PRB,Hohls2002PRL,Li2005PRL,Li2009PRL,amoo2014JPCM}. Theoretically, $z = 1$ is expected if the electrons in the IQHE regime are treated as \textit{interacting} particles with long-range (1/$r$) Coulomb interaction~\cite{Aleiner1994prb,hui2019prb,kumar2022prr}. On the other hand, in the \textit{non}-\textit{interacting} or screened (short-range interaction) limit, theory predicts exponents $z=2$ and $\kappa \simeq 0.21$ \cite{Aleiner1994prb, kumar2022prr}. An experimental confirmation of these exponents, however, has remained elusive for over three decades, until now.\\
\indent Here, we demonstrate an experimental platform where, by introducing a nearby \textit{screening} layer, we can directly tune the strength of electron-electron interaction at quantum criticality. While the use of a screening layer to modify electron-electron interaction, and thus many-body phenomena, has been reported in numerous material systems \cite{Aleiner1994prb, kumar2022prr, Pan1999prb, Deng2018prb, Goodwin2019prb, Pizarro2019prb, Stepanov2020Nature}, our experimental study applies this technique to the problem of quantum criticality and modification of critical exponents. Our platform is a $bilayer$ GaAs/AlAs/GaAs device, shown schematically in Fig.~\ref{fig:1}(a), consisting of two electron layers in close proximity. We leverage the fact that each electron layer undergoes transitions as a function of $B$, switching between strongly-insulating quantum Hall states, and conducting metallic states at the plateau-to-plateau transitions. We keep the bottom-layer density constant and measure $\kappa$ as this layer goes through a plateau-to-plateau transition at certain magnetic field. By tuning the electron density in the top layer, we can realize the following two scenarios:

(\textit{i}) The top layer is placed in a quantum Hall ($insulating$) state, resulting in an $unscreened$, long-range, Coulomb interaction between electrons in the bottom layer. Our data in this case yield $\kappa \simeq 0.42$, implying $z = 1$.

(\textit{ii}) The top layer is placed at a plateau-to-plateau transition so that it is in a metallic ($conducting$) state and therefore leads to a $screened$, short-range interaction between electrons in the bottom layer. In this case we find $\kappa \simeq 0.22$, consistent with $z = 2$. \\
\indent Our results provide direct experimental evidence for the role of interaction and its nature (long-range vs short-range) in quantum criticality in the IQHE regime.

Our device consists of two 2D electron systems in close proximity confined to two GaAs quantum wells (QWs) of width 18 nm, separated by a 5-nm-wide AlAs barrier, as shown in Fig.~\ref{fig:1}(a). The barrier is sufficiently wide to strongly suppress interlayer tunneling~\cite{footnote3}. This can be seen from the calculated electron charge distribution (Fig.~\ref{fig:1}(b)). The as-grown electron densities are $\simeq 5.5$ and $6.5$ (in units of $10^{10}$ cm$^{-2}$ which we use throughout this manuscript) in the top layer and bottom layer, respectively.  The mobility in the bottom layer is $ \simeq 1.3 \times 10^5$ cm$^2$V$^{-1}$s$^{-1} $. The device is an etched Hall bar of dimensions 200 $\mu$m $\times$ 2700 $\mu$m (Fig.~\ref{fig:1}(c)), with alloyed InSn contacts made to both electron layers. Four finger gates are patterned on top of the Hall bar arms, three of which (1, 2, and 3 in Fig.~\ref{fig:1}(c)) are energized to deplete the top layer so we can independently measure transport in the bottom layer~\cite{eisenstein1990apl}. Finger-gate 4 is electrically connected to the drain contact, and together they serve as the ground for the device. A global top gate ($V_{t}$) covers the active area of the Hall bar and is used to tune the density of top layer and therefore the magnetic fields at which its plateau-to-plateau transitions occur. The sample was cooled in a pumped $^{3}$He system and magnetoresistance measurements were carried out using standard lock-in techniques. The value of $\kappa$ was extracted through the temperature dependence of the maximum of the derivative of $R_{xy}$ with respect to $B$, which should follow the expression $dR_{xy}/dB|_{B = B_{c}} \propto T^{-\kappa}$~\cite{Pruisken1988PRL}.

\begin{figure}[]
\centering
\includegraphics[width=1.0\columnwidth]{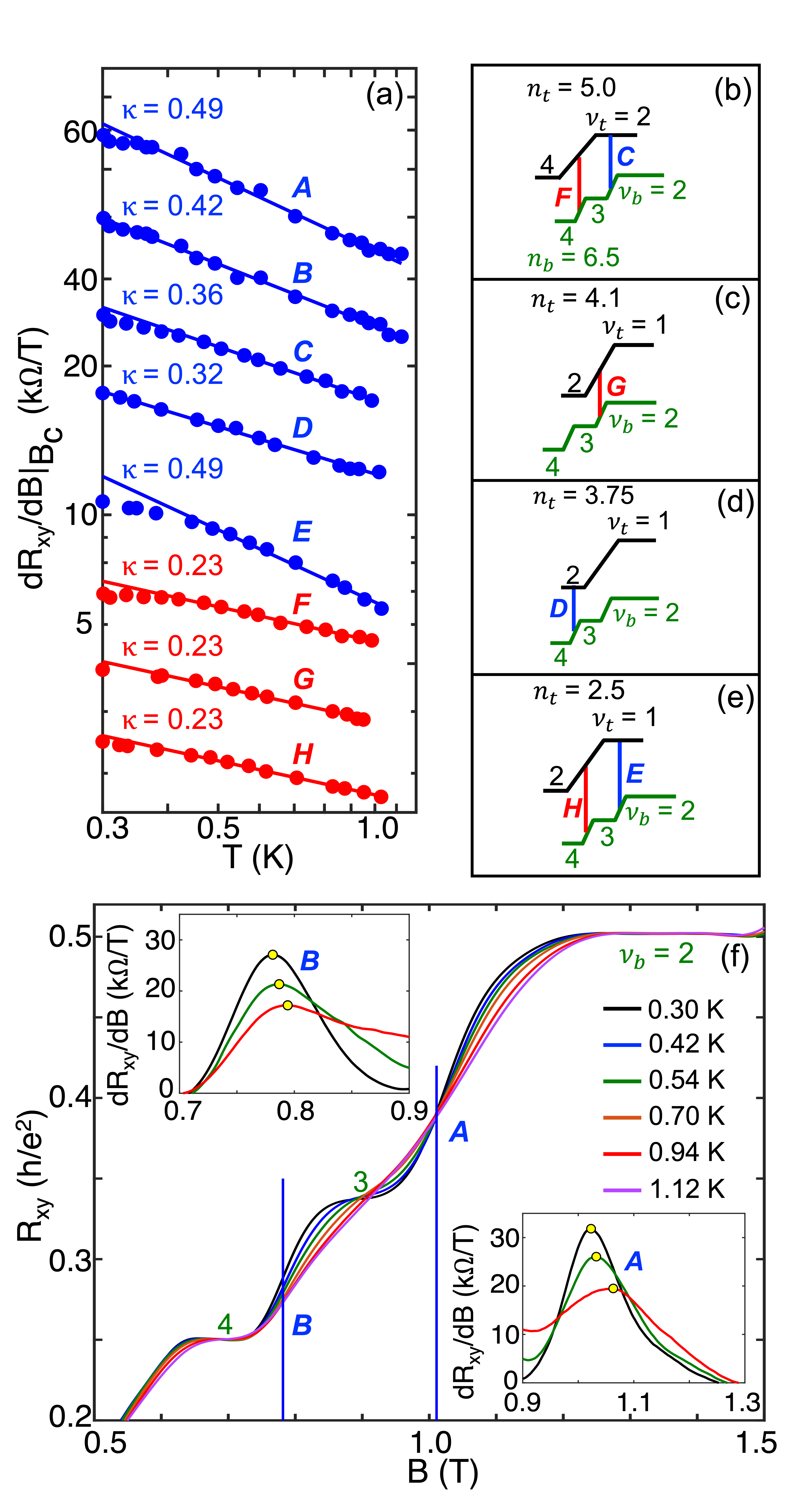}
\caption{ Summary of the measured $\kappa$ values in this work. (a) Temperature dependence of $dR_{xy}/dB|_{B = B_{c}}$ for different transitions, labeled \textbf{\textit{A}}-\textbf{\textit{H}}. Blue symbols are used for cases \textbf{\textit{A}}-\textbf{\textit{E}} where there is no screening provided by the top layer, and red symbols for cases \textbf{\textit{F}}-\textbf{\textit{H}} where there is screening. The data for each case are shifted vertically for clarity. Panels (b)-(e) show transitions \textbf{\textit{C}}-\textbf{\textit{H}} schematically. Black (green) curves represent the top-layer (bottom-layer) $R_{xy}$ traces, and $\nu_t$ and $\nu_b$ indicate their respective filling factors. Panel (f) presents the $R_{xy}$ traces for the bottom layer near integer quantum Hall plateaus $\nu_{b}=2$ to 4 as a function of temperature ranging from 0.30 K to 1.12 K when the top layer is completely depleted ($n_{t}=0$). Transitions \textbf{\textit{A}} and \textbf{\textit{B}} are marked. The insets show the derivative data ($dR_{xy}/dB$), at different temperatures at transitions \textbf{\textit{A}} and \textbf{\textit{B}}, and the values of $dR_{xy}/dB|_{B = B_{c}}$ are shown in Fig.~2(a).}
\label{log-log-and-Rxy}
\end{figure}

Figure~\ref{log-log-and-Rxy}(a) summarizes the extracted values of $\kappa$ at various electron densities of the top layer ($n_{t}$), while the density of the bottom layer ($n_{b}$) is kept at 6.5. Panel (a) shows $log$-$log$ plots of $dR_{xy}/dB|_{B = B_c}$ as a function of temperature for eight different cases, labeled \textbf{\textit{A}} to \textbf{\textit{H}}. In cases \textbf{\textit{A}} and \textbf{\textit{B}}, $n_{t}=0$, and in cases \textbf{\textit{C}} to \textbf{\textit{H}}, $n_{t}$ is varied from 5.0 to 2.5. The $R_{xy}$ traces for the top layer (black) and bottom layer (green) at different $n_{t}$ are schematically shown in panels (b)-(e).  For each case in Fig.~\ref{log-log-and-Rxy}(a), a line is fitted through the data points to determine the value of $\kappa$ from the expression $dR_{xy}/dB|_{B = B_c} \propto T^{-\kappa}$~\cite{footnote2, Kaur2024NC}; note that the data are vertically offset for clarity. We find that $\kappa \simeq 0.23$ for cases \textbf{\textit{F}}, \textbf{\textit{G}} and \textbf{\textit{H}}, much smaller than $\kappa$ for cases \textbf{\textit{A}} to \textbf{\textit{E}}.

To understand the data of Fig.~\ref{log-log-and-Rxy}(a), we first focus on two simple cases, \textbf{\textit{A}} and \textbf{\textit{B}}, where we used $V_{t}$ to completely deplete the electrons in the top layer ($n_{t} = 0$). The bottom-layer $R_{xy}$ traces under this condition are shown in Fig.~\ref{log-log-and-Rxy}(f) at temperatures ranging from 0.30 to 1.12 K. These traces capture the $\nu_{b} = 2$ to 3 and $\nu_{b}=3$ to 4 plateau-to-plateau transitions of the bottom layer (cases \textbf{\textit{A}} and \textbf{\textit{B}}). To analyze the data, we applied a Savitzky–Golay filter of order 2~\cite{savitzky1964anachem} to smooth both the raw $R_{xy}$ data and its derivative with respect to magnetic field. We then extracted the maximum of $dR_{xy}/dB$ from the smoothed data, as shown in Fig.~\ref{log-log-and-Rxy} insets for cases \textbf{\textit{A}} and \textbf{\textit{B}}. (For comparison of the smoothed and raw data, see Supplemental Material (SM)~\cite{Supp}.) As expected, the transitions become less sharp with increasing temperature, reflecting their thermal broadening, and the peak values in $dR_{xy}/dB$ decrease. These peak values are used to determine the scaling parameter $\kappa$, as shown in Fig.~\ref{log-log-and-Rxy}(a). The extracted $\kappa$ are 0.49 and 0.42 for transitions \textbf{\textit{A}} and \textbf{\textit{B}}, respectively. These values are reasonably consistent with previously reported values $ 0.3  \lesssim \kappa  \lesssim 0.6$ when there is no screening layer near the 2D electrons~\cite{wei1988prl, Li2005PRL,Koch1991PRB}.

\begin{figure*}[t]
\centering
\includegraphics[width=1\textwidth]{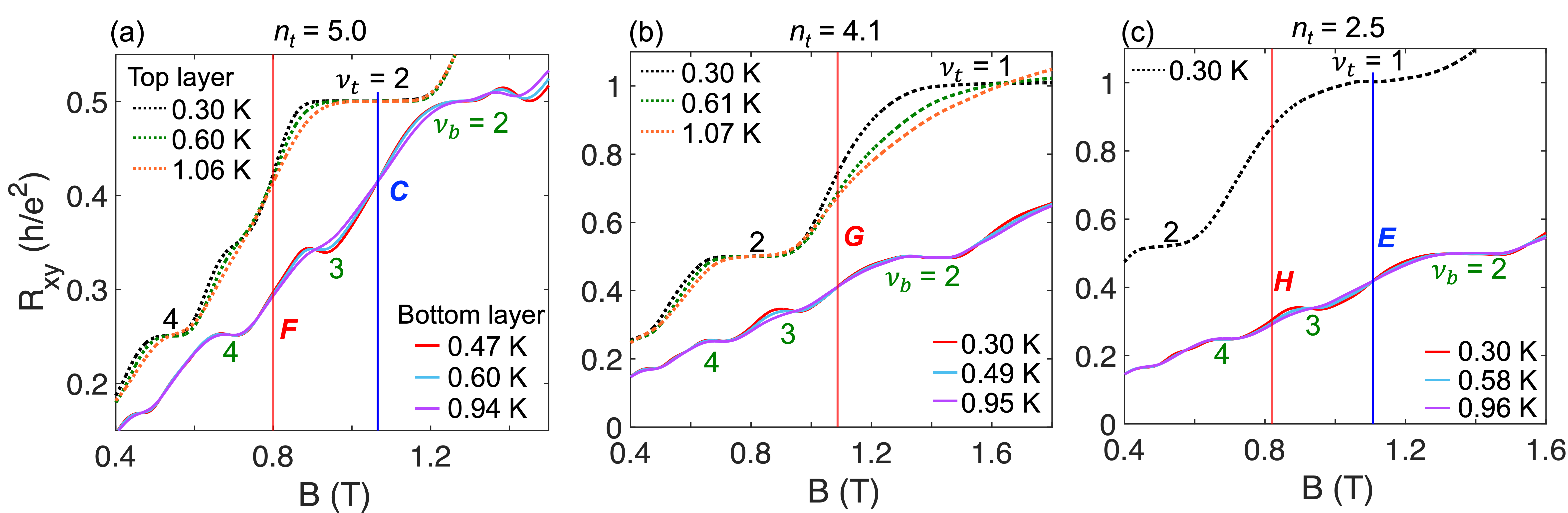}
\centering
\caption{Temperature dependence of $R_{xy}$ for the top layer (dotted traces) and bottom layer (solid traces). Panels (a)-(c) show the bottom layer plateau-to-plateau transitions at $n_t = 5.0$, 4.1, and 2.5, respectively, with the bottom-layer density fixed at $n_{b} = 6.5$.}
\label{Three-cases}
\end{figure*}

Next, we discuss scenarios where the top layer is populated with electrons ($n_{t} > 0$). We utilized the global top gate bias ($V_{t}$) to vary $n_{t}$, as indicated in Figs.~\ref{log-log-and-Rxy}(b-e), while the bottom layer density is kept fixed at $n_{b}$ = 6.5 (see SM~\cite{Supp} for the determination of the values of $n_{t}$ and $n_{b}$). The finger gates were activated, so that the transport measurements probed exclusively the bottom layer~\cite{footnote1,Supp}. We divide the six cases indicated in Figs.~\ref{log-log-and-Rxy}(b-e) into two categories. In cases \textbf{\textit{C}}, \textbf{\textit{D}}, and \textbf{\textit{E}}, the density in the top layer is tuned so that the plateau-to-plateau transitions in the bottom layer coincide with the top layer being in a quantum Hall state. As an example, $R_{xy}$ traces for case \textbf{\textit{C}} ($n_{t}$ = 5.0, $n_{b}$ = 6.5) are shown in Fig.~\ref{Three-cases}(a). Note that the $\nu_{b}$ = 2 to 3 plateau-to-plateau transition for the bottom layer matches the $\nu_{t}$ = 2 quantum Hall state of the top layer. The extracted $\kappa$ for this case \textbf{\textit{C}} is 0.36; see Fig.~\ref{log-log-and-Rxy}(a). As indicated in Fig.~\ref{log-log-and-Rxy}(a), the extracted $\kappa$ values for the other two cases in this category, \textbf{\textit{D}} and \textbf{\textit{E}}, are 0.32 and 0.49. These three values for cases \textbf{\textit{C}}, \textbf{\textit{D}}, and \textbf{\textit{E}} are all close to 0.42, and comparable to those for cases \textbf{\textit{A}} and \textbf{\textit{B}}, when $n_{t} = 0$. This is reasonable because in cases \textbf{\textit{C}}, \textbf{\textit{D}}, and \textbf{\textit{E}}, the top layer is insulating and cannot provide screening in the bottom layer. 

The highlights of our study are cases \textbf{\textit{F}}, \textbf{\textit{G}}, and \textbf{\textit{H}}, where the magnetic field positions of the quantum Hall plateau-to-plateau transitions in the top layer $and$ bottom layer match so that the top layer provides screening for the bottom layer. In Fig.~\ref{Three-cases}(b) $R_{xy}$ traces are shown for case \textbf{\textit{G}} ($n_{t}$ = 4.1, $n_{b}$ = 6.5), where the $\nu_{b}$ = 2 to 3 plateau-to-plateau transition for the bottom layer matches the $\nu_{t}$ = 1 to 2 plateau-to-plateau transition of the top layer. Similarly, in case \textbf{\textit{H}} ($n_{t}$ = 2.5, $n_{b}$ = 6.5), as shown in Fig.~\ref{Three-cases}(c), the $\nu_{b}$ = 3 to 4 plateau-to-plateau transition of the bottom layer matches the $\nu_{t}$ = 1 to 2 transition of the top layer. The traces for case \textbf{\textit{F}} ($n_{t}$ = 5.0, $n_{b}$ = 6.5) are shown in Fig.~\ref{Three-cases}(a). In this case, as the $R_{xy}$ traces for the top layer indicate, the $\nu_{t} = 3$ quantum Hall state, which corresponds to the Fermi level residing in a small, spin-split (Zeeman) energy gap, is not developed. The $R_{xy}$ traces for the top layer indicate a transition from $\nu_t$ = 2 to 4 while the bottom layer goes through its $\nu_{b}$ = 3 to 4 plateau-to-plateau transition. Remarkably, in all three \textbf{\textit{F}}, \textbf{\textit{G}}, and \textbf{\textit{H}} cases, where the top layer is in a conducting phase and provides screening in the bottom layer, the extracted values of $\kappa$ drop to 0.23, as shown in Fig.~\ref{log-log-and-Rxy}(a). \textit{It is evident that screening plays a crucial role and lowers $\kappa$ to 0.23}. 

\begin{figure}[h]
\centering
\includegraphics[width=0.8\columnwidth]{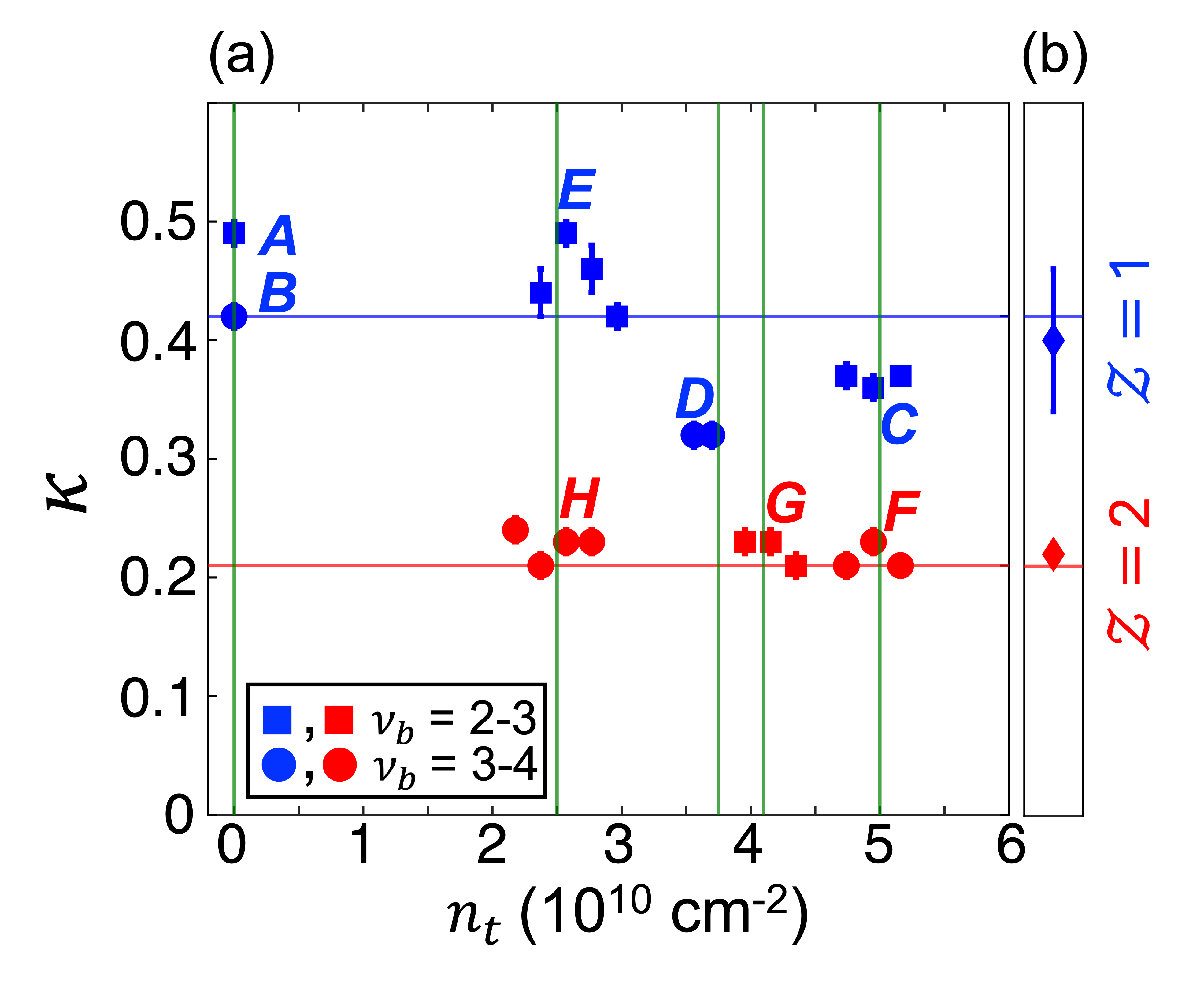}
\centering
  \caption{Measured $\kappa$ for the plateau-to-plateau transitions $\nu_{b}=2$ to 3 and 3 to 4 as a function of the top-layer density $n_t$. In panel (a), $\kappa$ are shown for cases \textbf{\textit{A}}--\textbf{\textit{H}}, with the corresponding $n_t$ values indicated by green vertical lines, together with additional data points in their vicinity, illustrating the evolution of $\kappa$ with $n_t$. Blue and red symbols represent the unscreened and screened, respectively. Square symbols correspond to the $\nu_{b} = 2$ to 3 transition, while circles represent the $\nu_{b} = 3$ to 4 transition. The horizontal lines indicate the theoretically expected universal values $\kappa = 0.42$ and $0.21$. Panel (b) shows the average of measured $\kappa$ for the unscreened (symbols) and screened (red symbols) cases, with error bars representing the standard deviation. Note that exponents $\kappa=0.42$ and $\kappa=0.21$ imply dynamic exponents $z=1$ and 2, as indicated on the right side of panel (b). 
  }
\label{kappa-density}
\end{figure}

Figure~\ref{kappa-density}(a) summarizes the extracted scaling exponent $\kappa$ as a function of $n_t$. Cases \textbf{\textit{A}}-\textbf{\textit{H}}, corresponding to the specific values of $n_t$ as quoted so far in the manuscript, are indicated by vertical green lines. We use blue symbols for the unscreened cases (\textbf{\textit{A}} to \textbf{\textit{E}}), and the red symbols for the screened cases (\textbf{\textit{F}} to \textbf{\textit{H}}). We also measured $\kappa$ for $n_{t}$ in the vicinity of each quoted case, and the results are also shown in Fig.~\ref{kappa-density}(a). The clustering of the measured $\kappa$ around each quoted case demonstrates the robustness of the exponents against small variations in $n_t$, and confirms the efficient screening ability, or lack of screening by the top layer. In Fig.~\ref{kappa-density}(b), we also show the averages of the $\kappa$ values of the blue and red data points. Two distinct regimes emerge: The unscreened (blue) data points, which have a large spread with an average value of $0.40\pm 0.06$, and the screened (red) data points which have a much narrower spread with an average value of $0.22\pm 0.01$. \\
\indent The above observations are consistent with theoretical expectations. Focusing on the red (screened) data points, we note that effective screening requires that the screening length scale ($d$), defined as the separation between the two layers ($\simeq 20$ nm in our sample, see Fig.~\ref{fig:1}), be smaller than or comparable to the magnetic length ($l_{B}=\sqrt{\hbar/eB}$). In our experiments, $d/l_{B} \simeq 1.0$ for the $\nu_{b}$ = 2 to 3 transition and $\simeq$ 0.8 for the $\nu_{b}$ = 3 to 4 transition, firmly establishing the screening capability of the top layer at these transitions. Another important point is that, in our sample, $d$ is much smaller than the distance between the layers and the $\delta$-doping layers ($\simeq 388$ nm), or the average distance between the residual impurities (estimated to be $\simeq 500$ nm), see~\cite{Chung2021NatMaterial, Chung2022prb}. This means that the screening layer in our sample screens both the electron-electron interaction within the bottom layer and the disorder potential experienced by this layer. Consequently, screening brings the bottom layer to the regime of short-range interaction \textit{and} short-range disorder, thus providing a natural explanation for our observation of $\kappa$ values which are very close to the universal value of $\kappa \simeq 0.21$, the exponent predicted for this regime \cite{Aleiner1994prb,kumar2022prr}.\\ 
\indent In contrast, the unscreened (blue) data points in Fig.~\ref{kappa-density}(a) have a large spread, and their average is $0.40 \pm 0.06$. Both the average value of $\simeq 0.40$ and the large spread are consistent with what has been previously reported for unscreened situations~\cite{wei1988prl,Koch1991PRB,Li2005PRL,Li2009PRL,Wei1992PRB,Gusev1998PRB,arapov2019LTP,amoo2014JPCM}. In these situations, the nature of the disorder potential plays a role, leading to variations and non-universality of the measured $\kappa$. For example, in standard (unscreened) 2D electron systems confined to modulation-doped GaAs/AlGaAs QWs, the dominant sources of disorder are charged impurities, leading to a long-range disorder potential, and varying $\kappa$ values~\cite{Wei1992PRB,Li2005PRL}. Li \textit{et al}.~\cite{Li2005PRL} managed to engineer a short-range disorder potential by intentionally introducing neutral Al dopants to the QW. They observed $\kappa \simeq 0.42$ with a narrow spread. Similarly, in trilayer graphene reported by Kaur \textit{et al.}~\cite{Kaur2024NC}, in the case of lower-quality devices dominated by long-range charged impurity disorder, the scaling exponent deviated from a universal value. However, when the disorder potential was effectively screened using a nearby metallic graphite gate~\cite{footnote4}, the exponent $\kappa \simeq 0.42$ was recovered, consistent with short-range disorder. Hence, the non-universality of $\kappa$ observed in the unscreened cases of our experiments is consistent with what has been reported in the literature.\\ 
\indent Returning to Fig.~\ref{kappa-density}(b), it is remarkable that the screened (red) data points exhibit an exceptionally narrow distribution, and have an average value of $0.22\pm 0.01$. These observations are both consistent with theoretical expectations. The reduced spread in $\kappa$ reflects the screening action of the top layer which makes the criticality less sensitive to the exact nature of the disorder potential. More importantly, the effective screening by the electrons in the top layer leads to a short-range interaction of the electrons in the bottom layer, which is essential for observing a universal value of $\kappa = 0.21$ ($z = 2$), as theoretically predicted~\cite{Aleiner1994prb, kumar2022prr}. Our results therefore validate the theoretical results and demonstrate how quantum criticality can be tuned \textit{in-situ} in a single device. \\
\indent We close with two remarks. First, while the IQHE provides a unique platform to study the interplay between localization and interaction, the critical exponents we report here are not specific to IQHE, but also pertain to a broader universality class. In particular, the scaling behavior in the IQHE is widely understood in terms of quantum percolation physics, as captured by the Chalker–Coddington network model~\cite{Chalker1988jpc}. Within this framework, the plateau transition arises from percolation of equipotential contours combined with quantum tunneling and interference at saddle points. Our results therefore directly address the role of electron–electron interactions at criticality in such quantum percolation networks. The observation of a reduced exponent $\kappa \simeq 0.22$ in our system, where interactions are effectively screened, provides strong evidence that interactions can modify the universality class of the transition. This makes our findings broadly relevant, particularly in the context of understanding localization and interaction effects in low-dimensional systems.\\
\indent Second, our experimental platform can be naturally extended to the strongly interacting fractional quantum Hall effect (FQHE) regime. Superuniversality arguments predict that the same scaling laws govern both IQHE and FQHE transitions in the presence of long-range Coulomb interaction~\cite{kumar2022prr,Pu2022prl}. Consistent with this expectation, experiments in the FQHE regime generally report $\kappa \simeq 0.42$ with a reasonable spread in the data points~\cite{Engel1990SS,Madathil2023PRL,Kaur2024NC}. The fate of superuniversality in the screened (short-range interaction) limit remains unresolved, even theoretically. An extension of experiments such as ours to the FQHE regime can offer a promising route to elucidate the nature of quantum critical scaling in this unexplored regime. Such studies present additional challenges. In particular, achieving the regime $d/l_{B} \lesssim 1$, where interaction can be strongly modified, typically requires access to lower carrier densities because of the low filling factors involved in realizing FQHE. At such low densities, disorder-induced localization and reduced many-body gaps can obscure the emergence of robust FQHE states. Nonetheless, recent advances in ultra-high-quality GaAs heterostructures at very low carrier densities~\cite{Chung2022prl}, as well as very high-quality, atomically-thin 2D materials~\cite{Dean2020book,Domaretskiy2025nature} provide a promising route to explore these questions. These developments open the possibility of testing interaction-driven modifications to superuniversality, in the FQHE regime.

\begin{acknowledgments}
We acknowledge support by the National Science Foundation (NSF) Grants DMR-2104771 and DMR-2611783 for measurements, and by the Gordon and Betty Moore Foundation's EPiQS Initiative (Grant No.~GBMF9615.01 to L.~N.~P.) for sample fabrication. We also thank Jainendra K. Jain and Prashant Kumar for illuminating discussions.
\end{acknowledgments}



\begin{thebibliography}{99}
\bibitem{footnote3}
The calculated separation between the symmetric and antisymmetric energy levels, which is a measure of interlayer tunneling in our structure, is 0.06 K.
\end{thebibliography}
\end{document}


\setcounter{page}{1}

\title[]{Supplemental Material to "Engineering Quantum Criticality in the Integer Quantum Hall Regime through a Screening Layer"}
\author{C.~T.~Tai}
\thanks{These authors contributed equally to this work.}
\author{P.~T.~Madathil}
\thanks{These authors contributed equally to this work.}
\author{A.~Gupta}
\author{L.~N.~Pfeiffer}
\author{K.~W.~Baldwin}
\author{M.~Shayegan}

\affiliation{Department of Electrical Engineering, Princeton University,
Princeton, New Jersey 08544, USA}

\date{\today}
\maketitle


\section*{\RNum{1}.\indent Details of Savitzky Golay smoothing process}

\begin{figure}[h]
\centering
\includegraphics[width=1\textwidth]{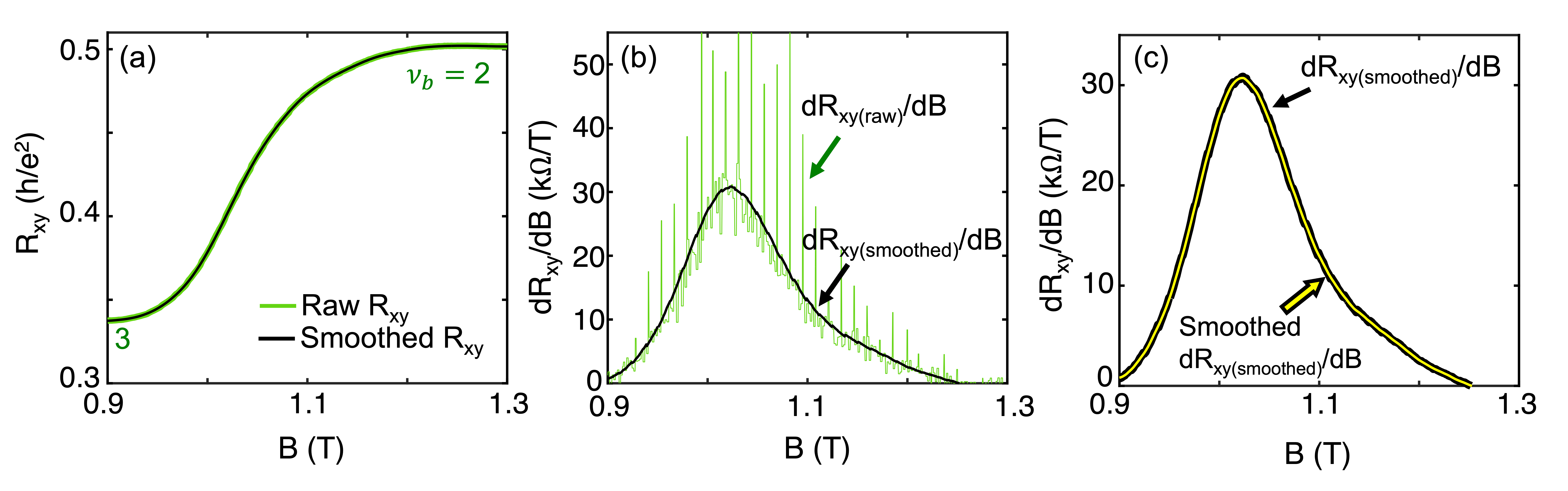}
\caption{ Panel (a) shows the smoothed $R_{xy}$ data at $T = 0.30$ K (black), overlaid on the raw $R_{xy}$ trace (green). Panel (b) presents the corresponding $dR_{xy}/dB$ for each case: the green curve represents the derivative of raw $R_{xy}$ ($dR_{xy\text{(raw)}}/dB$), while the black curve shows the derivative of the smoothed $R_{xy}$ ($dR_{xy\text{(smoothed)}}/dB$). Lastly, the Savitzky–Golay filter is applied to the $dR_{xy\text{(smoothed)}}/dB$. The $dR_{xy\text{(smoothed)}}/dB$ and the smoothed $dR_{xy\text{(smoothed)}}/dB$ are shown in panel (c).
}
\label{fig:filter}
\end{figure}

In this section, we discuss the filtering procedure we use to obtain $dR_{xy}/dB$ values by presenting, as an example, case \textbf{A} of the main text (Fig.~2(a)) at \textit{T} = 0.30 K in Fig. S1. The raw $R_{xy}$ traces were first smoothed using a Savitzky–Golay filter of order 2 and a window size of 0.1 T. Figure~\ref{fig:filter}(a) shows the smoothed $R_{xy}$ data (black) overlaid on the raw trace (green) at $T = 0.30$ K in case \textbf{\textit{A}}. For comparison, we show the derivative of the raw $R_{xy}$ [$dR_{xy\text{(raw)}}/dB$, showed in green] together with the derivative of the smoothed $R_{xy}$ [$dR_{xy\text{(smoothed)}}/dB$, showed in black] in Fig.~\ref{fig:filter}(b). It is clear that $dR_{xy\text{(smoothed)}}/dB$ is significantly cleaner and preserves the underlying features. Finally, the Savitzky–Golay filter is applied again to the $dR_{xy\text{(smoothed)}}/dB$ trace. A comparison between $dR_{xy\text{(smoothed)}}/dB$ and the smoothed $dR_{xy\text{(smoothed)}}/dB$ is presented in Fig.~\ref{fig:filter}(c) as black and yellow curves, respectively. The filter order is 2 and window size is 0.02 T. The $\kappa$ values reported in our study are based on the smoothed $dR_{xy\text{(smoothed)}}/dB$ curves. As seen in Fig.~\ref{fig:filter}, the smoothed $R_{xy}$ and $dR_{xy}/dB$ are in excellent agreement with the corresponding raw data, and no artifacts are introduced in our smoothing procedure.  

\clearpage

\section*{\RNum{2}.\indent  Determination of Top and Bottom Quantum Well Densities}

\begin{figure}
\centering
\includegraphics[width=1\textwidth]{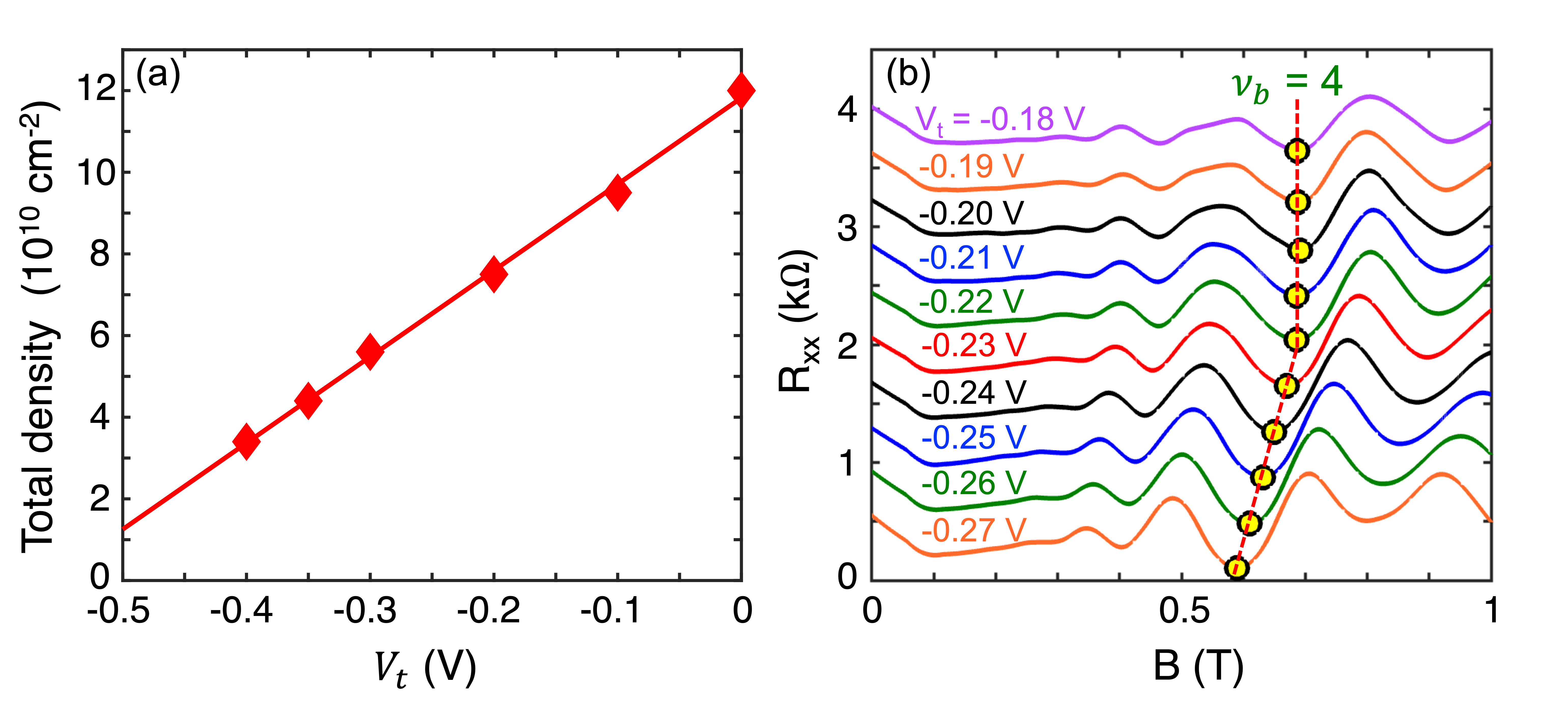}
\caption{(a) Total electron density vs $V_t$. Densities are extracted from the slope of the Hall data, with all the finger gates grounded. A linear fit is applied to calculate the gating efficiency. (b) $R_{xx}$ traces at different $V_t$ values ranging from $-0.18~\mathrm{V}$ to $-0.27~\mathrm{V}$ with the finger gates activated, so that only the bottom layer is measured. The yellow circles mark the $\nu_{b} = 4$ minima and the red dashed line tracks their positions. The shift of the $\nu_{b} = 4$ minimum at $V_t = -0.23~\mathrm{V}$ indicates complete depletion of the top layer, from which we can calculate the as-grown density of the bottom layer. The traces are offset by $400~\Omega$ for better visibility.}
\label{fig:S1}
\end{figure}

Figure~\ref{fig:S1} summarizes how we determine the top-layer density ($n_{t}$) and the gating efficiency in our sample. To tune $n_{t}$ to a desired value, we have to know the as-grown density of the top layer and its gating efficiency. In Fig.~\ref{fig:S1}(a), we extract the total electron density from Hall measurements without energizing the finger gates; therefore, the measured density corresponds to the total density $(n_{t}+n_{b})$. Because the interlayer separation ($\simeq 23$ nm) is only $\simeq 3\%$ of the distance from the surface to the center of the top layer ($\simeq 734$ nm), we assume that the difference in gating efficiency between the two layers is negligible. This assumption is supported by the fact that there is no kink in Fig.~\ref{fig:S1}(a) and the six data points can be well fitted by a single straight line, yielding a gating efficiency of $2.06 \times 10^{8}$~cm$^{-2}$/mV. \\
\indent The as-grown density of the bottom layer is determined by following the field positions of $\nu_{b}=4$ minima at various $V_{t}$ values, as shown in Fig.~\ref{fig:S1}(b). The field positions of $\nu_{b}=4$ minima, marked by yellow circles, remain fixed for $-0.22~\mathrm{V} \leq V_{t} \leq -0.18~\mathrm{V}$, indicating that $n_{t}$ is still finite and screens the applied gate bias. A clear shift of the $\nu_{b}=4$ minimum emerges at $V_{t}=-0.23$ V (red trace), signaling full depletion of the top layer. Starting from this point, for more negative $V_{t}$, $n_{b}$ decreases in accordance with the gating efficiency obtained in Fig.~\ref{fig:S1}(a). Based on the $\nu_{b}=4$ position at $V_{t}=-0.23$ V, we deduce an as-grown $n_{b} \simeq 6.5 \times 10^{10}$~cm$^{-2}$. Combining this with the total density measured at $V_{t}=0$ (Figure~\ref{fig:S1}(a)), we obtain an as-grown $n_{t} \simeq 5.5 \times 10^{10}$~cm$^{-2}$. Knowing the as-grown densities and gating efficiency, we can then tune $n_{t}$ or $n_{b}$ to the desired values.

\clearpage

\section*{\RNum{3}.\indent $\bm{dR_{xy}/dB}$ vs $\bm{B}$ traces for different cases}

\begin{figure}[h]
\centering
\includegraphics[width=1\textwidth]{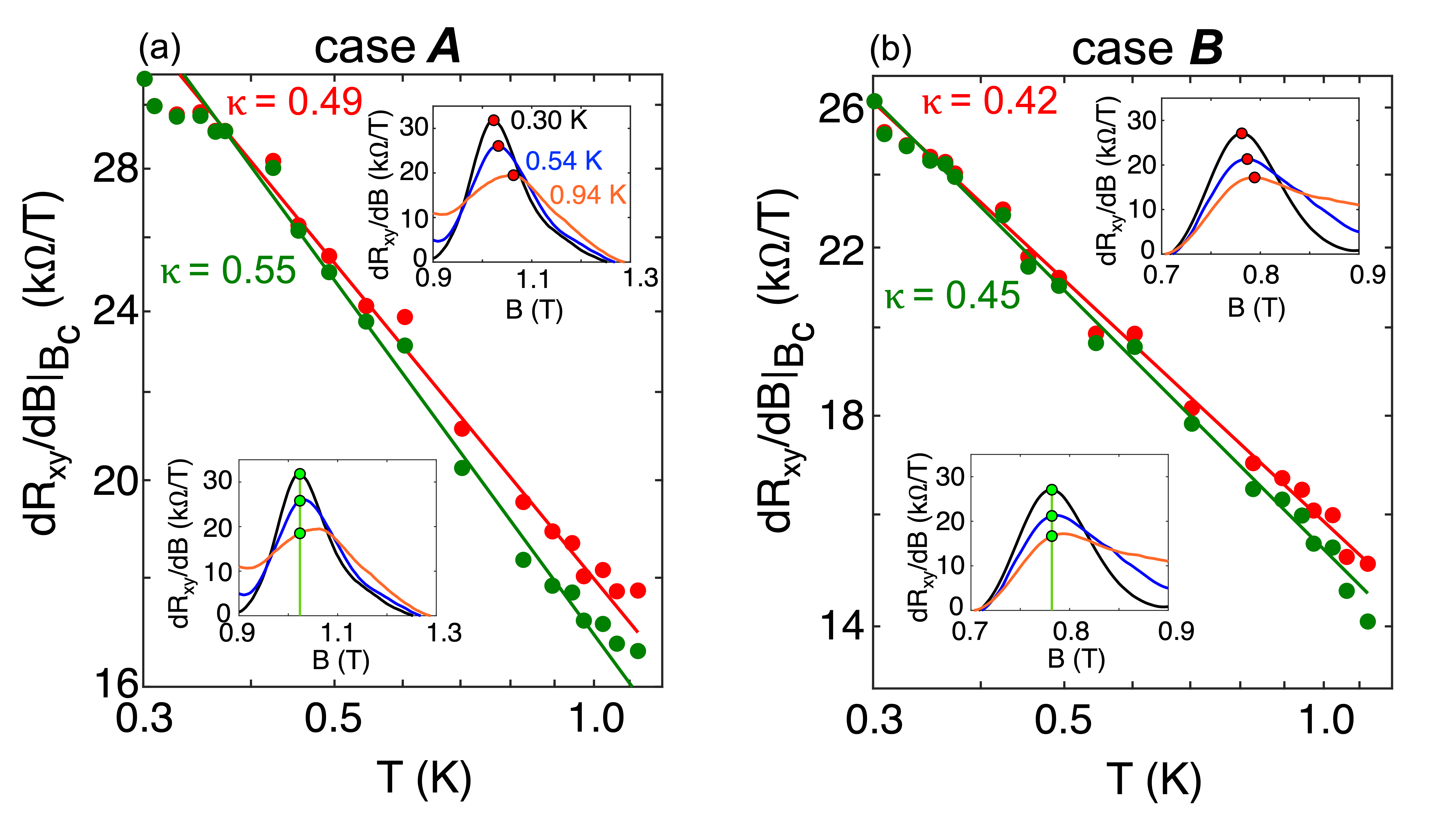}
\caption{Temperature dependence of the critical magnetic field $B_{c}$ for cases \textbf{\textit{A}} (left) and \textbf{\textit{B}} (right). Insets show three representative $dR_{xy}/dB$ traces at temperatures 0.30, 0.54, and 0.94 K, showing the shift in the peak position. Top insets in (a) and (b): red circles mark the $dR_{xy}/dB$ maxima, and the red lines in the main figures are power law fits used to extract $\kappa$ (first method). Bottom insets: green circles denote a fixed ($T$-independent) $B_{c}$ defined at 0.30 K. The green lines in main figures are power law fits used to extract $\kappa$ (the second method).
}
\label{fig:A and B}
\end{figure}

\begin{figure}[h]
\centering
\includegraphics[width=1\textwidth]{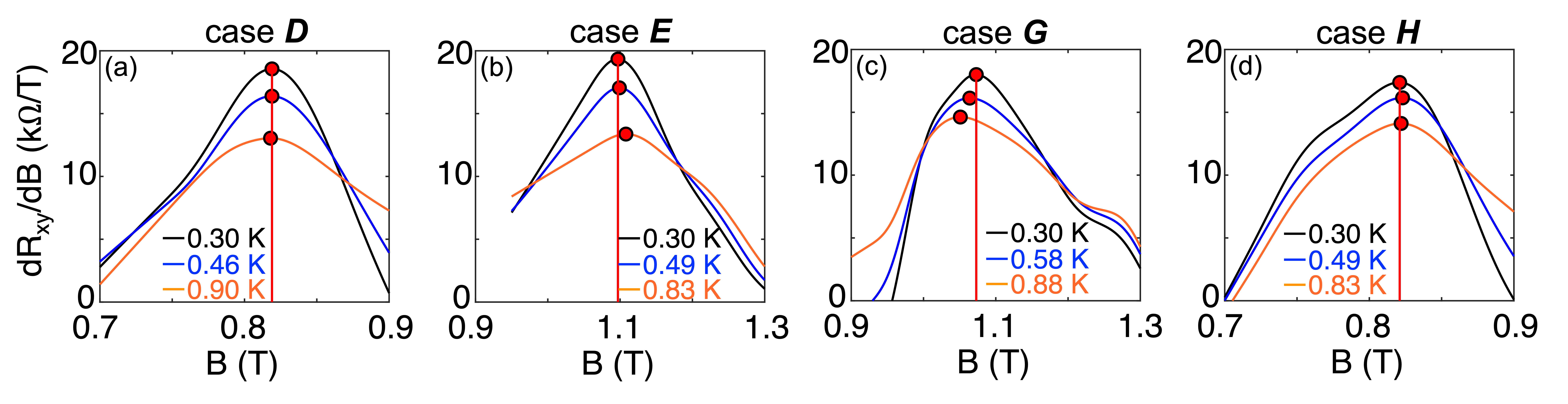}
\caption{Three representative $dR_{xy}/dB$ traces are shown at temperatures of $0.30$ K, $\simeq 0.5$ K, and $\simeq 0.9$ K for each of \textbf{\textit{D}}, \textbf{\textit{E}}, \textbf{\textit{G}}, and \textbf{\textit{H}} cases. 
The red circles mark the maxima of the derivative at different temperatures, and the vertical red lines indicate the positions of $B_{c}$ defined at 0.30 K.
}
\label{fig:E to H}
\end{figure}

As seen in insets to Fig. 2(f) of the main text, the positions of $dR_{xy}/dB$ maxima change slightly with temperature, implying that there is no fixed ($T$-indepenedent) critical magnetic field. Here we examine how this change affects the extracted critical exponent $\kappa$. Data for cases \textbf{\textit{A}} and \textbf{\textit{B}} are summarized in Fig.~\ref{fig:A and B}.  In the insets of Fig.~\ref{fig:A and B}, we reproduce the $dR_{xy}/dB$ vs $B$ traces presented in Fig. 2(f) insets. In the upper insets in Fig.~\ref{fig:A and B}, the peaks in the $dR_{xy}/dB$ traces, corresponding to the positions of $B_{c}$, are marked by red circles. We note that the positions of the maxima change by $\simeq4\%$ as temperature is changed from 0.30 K to 0.94 K. In the lower insets, we highlight the low temperature positions of $B_{c}$ with the green vertical lines. Next, we extract $\kappa$ from the following two methods. First, we assume that $B_{c}$ corresponds to the maximum of $dR_{xy}/dB$ at a given temperature, allowing for a varying $B_{c}$. The red data points in Fig.~\ref{fig:A and B} show the $dR_{xy}/dB|_{Bc}$ vs $T$, and fits through these points yield $\kappa$ values of 0.49 and 0.42 for cases  \textbf{\textit{A}} and \textbf{\textit{B}}, respectively. These are the values reported in the main text. Second, we assume a fixed $B_{c}$ corresponding to the low temperature limit and extract $dR_{xy}/dB|_{Bc}$ as shown in the lower insets of Fig.~\ref{fig:A and B}. The extracted $\kappa$ for this method, shown by the fits through the green data points are 0.55 and 0.45, respectively. The maximum variation in $\kappa$ is $\simeq12\%$, which is reasonable and consistent with our observed range of $0.3 \lesssim \kappa \lesssim 0.6$ in the absence of screening (see Fig. 4 of main text).

Next, we show representative $dR_{xy}/dB$ vs $B$ traces for cases \textbf{\textit{D}}, \textbf{\textit{E}}, \textbf{\textit{G}}, and \textbf{\textit{H}} in Fig.~\ref{fig:E to H}. The peaks in the $dR_{xy}/dB$ traces, corresponding to the positions of $B_{c}$, are marked by red circles. We also highlight the low temperature positions of $B_{c}$ with the red lines. There is a maximum of $2\%$ variation in $B_{c}$ (case \textbf{\textit{G}}). Our analysis of these data using the two methods described in the preceding paragraph yields $\kappa$ which have the same values to within $\simeq9\%$.

\begin{figure}
\centering
\includegraphics[width=1\textwidth]{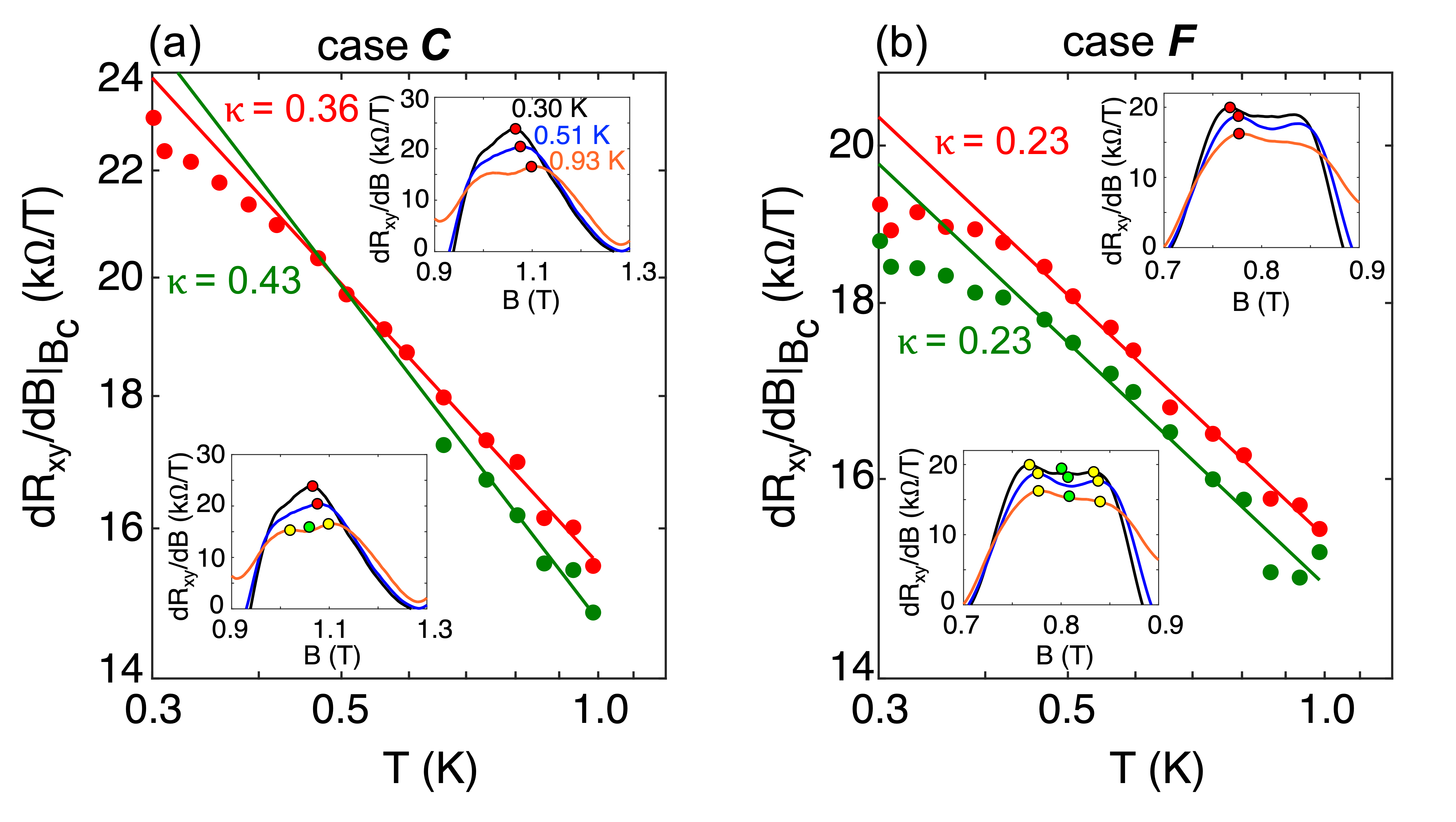}
\caption{
$dR_{xy}/dB|_{B_{c}}$ vs $T$ for cases \textbf{\textit{C}} (left) and \textbf{\textit{F}} (right). Insets display three representative $dR_{xy}/dB$ data at 0.30 K, 0.51 K, and 0.93 K. In the top insets in (a) and (b), red circles mark the maximum $dR_{xy}/dB$ values, and the red lines in main figures are power law fits used to extract $\kappa$ (first method). In the bottom insets, for traces where $dR_{xy}/dB$ exhibits two maxima, the two peaks are marked by yellow circles and their average value by the green circles. (For case \textbf{\textit{C}}, the $dR_{xy}/dB$ traces exhibit two maxima only for temperatures $\gtrsim 0.65$ K. For lower temperatures, there is a well-defined maximum.) The corresponding fits are shown as the green lines in the main figures. 
}
\label{fig:C and F}
\end{figure}

Finally, we discuss cases \textbf{\textit{C}} and \textbf{\textit{F}}. As shown in the insets, we observe broad or double-maxima in the $dR_{xy}/dB$ vs $B$ traces. Here again, we present two methods for extracting $\kappa$. First, we assume $B_{c}$ to correspond to the maximum of $dR_{xy}/dB$ and proceed to obtain $\kappa$, as shown by the red data points and fits in Fig.~\ref{fig:C and F}. We find $\kappa \simeq 0.36$ and 0.23 for cases \textbf{\textit{C}} and \textbf{\textit{F}}, respectively. Second, in traces where the maximum is broad or splits into two peaks, we perform the analysis by assuming that $B_{c}$ is the average of the two local maxima (shown in yellow in the lower two insets of Fig.~\ref{fig:C and F}), and extract $dR_{xy}/dB|_{B_{c}}$, as marked by the green circles. The $\kappa$ values obtained using this method are 0.43 and 0.23. We note the reasonable agreement in the $\kappa$ values determined from the two methods.


    